\documentclass[prl,aps,twocolumn]{revtex4}
\usepackage{graphicx}
\begin{document}
\title{Weber blockade in superconducting nanowires}
\author{Tyler Morgan-Wall}
\author{Benjamin Leith}
\author{Nikolaus Hartman}
\author{Atikur Rahman}
\author{Nina Markovi\'c}
\affiliation{{\it Department of Physics and Astronomy, Johns Hopkins
University, Baltimore, Maryland 21218, USA.}}
\begin{abstract}
We have measured the critical current as a function of magnetic field in short and narrow superconducting aluminum nanowires. In the range of magnetic fields in which vortices can enter a nanowire in a single row, we find regular oscillations of the critical current as a function of magnetic field. The oscillations are found to correspond to adding a single vortex to the nanowire, with the number of vortices on the nanowire staying constant within each period of the oscillation. This effect can be thought of as a Weber blockade, and the nanowires act as quantum dots for vortices, analogous to the Coulomb blockade for electrons in quantum dots.
 
\end{abstract}
\pacs{73.20.Mf, 73.20.Qt, 73.61.Ph, 73.63.-b}
\maketitle
Vortices in superconductors are topological excitations that carry a single flux quantum of $\Phi_{0} =h/2e$ and can be viewed as basic degrees of freedom that describe the low-energy states of the system \cite{Tinkham}. The shape of the vortices, their interactions, and the configurations in which they can exist in a superconductor are strongly affected by the dimensions and the geometry of the sample. In thin films, in which the coherence length $\xi$ is larger than the film thickness d, the vortices are of the Pearl type \cite{Pearl}, and are shaped like pancakes. As the width of the film is reduced, the vortices in narrow strips arrange themselves in rows \cite{Vodolazov}. Vortices cannot exist in strips or nanowires that are narrower than $\xi$, and such nanowires will eventually cross over to an insulating regime upon further width reduction\cite{Markovic}. However, if the width of a nanowire is on the order of a few $\xi$, the vortices will be able to enter the nanowire in a single row. If such a nanowire is also short enough that the energy difference between N and N+1 vortices is larger than kT, it may behave as a zero-dimensional quantum dot for vortices. Analogous to the phenomenon of Coulomb blockade for electrons in quantum dots, short and narrow strips could exhibit Weber blockade for vortices. This type of system was studied theoretically by Pekker \textit{et al.}, \cite{Weber} who used the charge-vortex duality \cite{MPAF} and the formalism developed for quantum dots \cite{Beenakker} to construct a model for the Weber blockade in a superconducting nanowire. A superconducting nanowire was also modeled as a line junction of Josephson vortices \cite{Efrat1, Efrat2}, motivated by the experimentaly observed periodic oscillations in magnetoresistance \cite{Shahar}.
In this work, we describe an experimental realization of a Weber-blockaded vortex dot in short superconducting aluminum nanowires. We find well defined diamond-shaped regions of zero resistance as a function of the current and magnetic field (vortex conductivity is zero in those regions, in analogy with the zero conductivity for electrons in Coulomb blockade). These Weber diamonds repeat with a periodicity that corresponds to adding a single flux quantum to the nanowire.

The aluminum nanowires were patterned in a four probe geometry using electron beam lithography. Cold developer method  \cite{coldmibk} was used to improve the resolution of the fabrication process, yielding 50nm wide nanowires. 25nm of aluminum was deposited in a thermal evaporator in a vacuum of $10^{-7}$ Torr at a rate of ∼1.5nm/s. The samples were placed in an Oxford 3He cryostat and cooled down to $250mK$. Four probe measurements of current-voltage characteristics were obtained by sourcing the current using a Keithley 6220 precision current source, and measuring the voltage using a Keithley 2182A nanovoltmeter. Magnetic field was applied in a direction perpendicular to the plane of the substrate.  

\begin{figure}
  \includegraphics[width=8 cm]{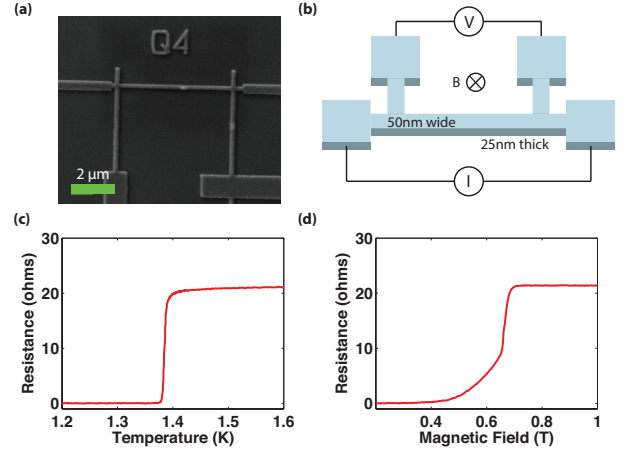}
  \caption{{\bf (a)} Scanning electron microscope image
  of an aluminum nanowire. The scale bar is 2 $\mu$m long.
  {\bf(b)} Schematic of the sample and the measurement. The measured nanowires are 25nm thick, 50-70nm wide and range 
  from 1.5-4.5$\mu$m in length between the current leads. The resistance measurements are carried out in a four-probe configuration, and the magnetic field is applied in a direction perpendicular to the plane of the substrate. {\bf(c)} Resistance as a function of temperature for one of
  the 1.5$\mu$m long nanowires. The normal state resistance $R_{N}$ is $20 \Omega$ and the $T_{C}$ is 1.38K. {\bf(d)} Resistance as a function of magnetic field for the same nanowire as in c). The normal state resistance is reached around $0.63T$.}
  \label{1}
\end{figure}

A scanning electron microscope image of an aluminum nanowire is shown in Fig. 1.a. and Fig. 1. b. shows the schematic of the nanowire and the measurement. The nanowires are 50-70nm wide, 25nm thick, 50nm wide and range
from 1.5-4.5$\mu$m in length. They are contacted by aluminum leads, and the width of the voltage leads is the same as the width of the nanowire at the point of contact. Based on the upper critical field measurements on films that were fabricated under the same conditions, the coherence length is estimated to be around $\xi$=$27nm$. This is about the same as the thickness, but smaller than the width of the wire. The resistance as a function of temperature in zero magnetic field for one of the the 1.5$\mu$m long nanowires is shown in Fig. 1.c. The superconducting transition occurs at 1.4K, and the normal state resistance is $20\Omega$. The resistance as a function of magnetic field at $250mK$ is shown in Fig. 1. d. Normal state resistance is reached at around $0.63T$. We will focus on the superconducting regime (at $250mK$ and below $0.4T$ in Fig. 1. d.) and study the magnetic field dependence of the critical current required for the onset of the resistive state.

\begin{figure}
    \includegraphics[width=8 cm]{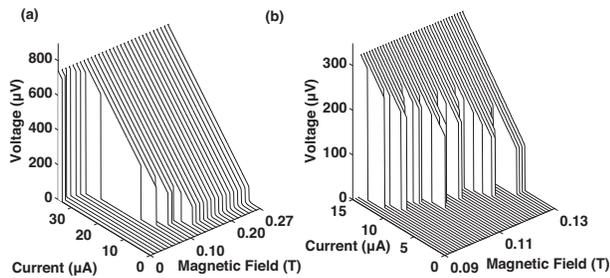}
  \caption{{\bf(a)} Current-voltage characteristics as a function of magnetic field from 0 to 270mT for a 1.5$\mu$m long nanowire at 250mK.  {\bf(c)} Current-voltage characteristics for the same nanowire as in (a), zoomed in the range of magnetic fields between 90 and 130mT, where the critical current shows non-monotonic behavior as a function of magnetic field.}
  \label{2}
\end{figure}

The current-voltage characteristics in magnetic fields from 0 to 270mT for a 1.5$\mu$m long nanowire at 250mK are shown in Fig 2.a. At low fields and low currents, the nanowire is in the superconducting state and the measured voltage is zero. When a current of about 34$\mu$A is reached, a sharp transition to the normal state is observed, with the voltage rising from zero to the normal state value within 50-100nA. We take the current at which this sharp transition occurs to be the critical current $I_{C}$ (as defined here, $I_{C}$ signifies the onset of non-zero resistance and is generally lower than the depairing current at which superconductivity is destroyed \cite{Tinkham}). When the magnetic field reaches about 60mT, the transition remains sharp, but $I_{C}$ drops to around 10$\mu$A. As the magnetic field is increased further, $I_{C}$ begins to oscillate, with several peaks and dips shown in a close-up in Fig. 2.b.  

The critical current as a function of magnetic field in the entire superconducting regime for same nanowire is shown in Fig. 3. a. After the initial drop around 60mT, oscillations in the critical current are clearly visible in the range of magnetic fields between 80-127mT. When the magnetic field reaches 127mT, the oscillations stop (rather suddenly), and the critical current decreases linearly to zero. A color plot of the voltage as a function of magnetic field and the applied current is shown in Fig. 3. b., taken with a 1mT step size in magnetic field. In order to distinguish these oscillations from the gaussian fluctuations of the critical current \cite{Tinkham}, we also carrried out a high-resolution magnetic field and current scan of the voltage multiple times, and averaged the resulting data (see Supplemental material for an animated image of the successive measurements \cite{Supp}). Figure 3.c. shows an average over seven scans, taken with the step size of 0.2mT. There is a clear pattern in the critical current as a function of the magnetic field (represented by the boundary between the black and white areas in Fig. 3. c.). Apart from a few small irregularities, the critical current increases linearly, reaches a peak, and then decreases linearly, with the average slopes shown in Fig. 3. a. This pattern repeats with a periodicity of about 5mT, and it stops rather abruptly upon reaching a peak at 127mT. Similar behavior was observed in nanowires that were up to 4.5$\mu$m long, as measured between the edges of the current leads (see Supplemental Material for data on additional samples \cite{Supp}). 

\begin{figure}
  \includegraphics[width=8 cm]{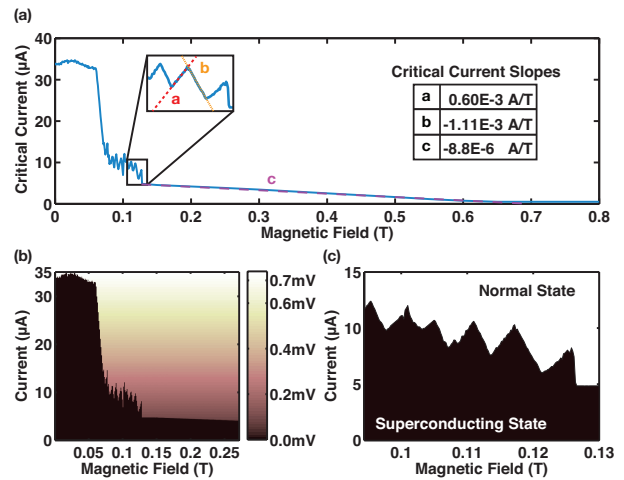}
  \caption{{\bf(a)} Critical current as a function of magnetic field for a 1.5$\mu$m long wire at 250mK, with a close-up of the oscillations. The avearge slopes of three linear regions are listed in the inset. 
  {\bf(b)} A color plot of the voltage as a function of magnetic field and current at 250mK. The data were taken during a single magnetic field sweep with 1mT step size. The dark areas are superconducting, and the voltage there is zero.
  {\bf(c)} A color plot of the voltage as a function of magnetic field and current at 250mK, in the range of fields between  90mT and 130mT. The data were taken using a magnetic field step size of 0.2mT and were averaged over seven magnetic field sweeps.}
  \label{3}
\end{figure}

As we argue below, the critical current oscillations can be understood in terms of discrete entry of vortices in the nanowire.
The appearance of vortices in thin films of superconductors has been studied extensively, both theoretically \cite{Likharev, Clem, Maksimova, Sanchez, Bronson, Palacios, Berdiyorov, Vodolazov, Edges, Aranson, Kuznetsov} and experimentally \cite{Stan, Plourde, Gutierrez}. In order to understand the observed oscillations of the critical current in magnetic field, we have to consider the characteristic dimensions of our samples. In thin and narrow strips, in which both the thickness and the width are of the order of $\xi$, the finite size of the vortex core cannot be ignored and one should use the Ginzburg-Landau model \cite{Aranson, Palacios, Sanchez, Maksimova}, rather than the London theory. Within the Ginzburg-Landau model, it has been argued that vortices can enter a thin film strip if its width is at least 1.8$\xi$ \cite{Vodolazov}. Assuming that the coherence length in our samples is about $\xi$=27nm, the width of our samples (50nm) is just large enough to allow entry of a single row of vortices. 
In general terms, the stability of vortices in a superconducting strip is governed by their potential energy (Gibbs free energy), which varies accross the width of the strip \cite{Weber, Sanchez, Stejic} as shown in Figure 4. a. At low magnetic fields, the potential energy is positive everywhere, and the vortices cannot enter the sample. 
As the magnetic field is increased to $B_{0}$, the potential energy starts to develop a dip in the center of the film, but it remains positive everywhere. Under these conditions, the vortices can exist in the film in a metastable state, trapped by the potential barriers at the edges of the strip \cite{Clem} (either entering by thermal activation, or remaining trapped after the magnetic field has been decreased from a higher value). Upon further increase of the magnetic field to $B_{S}$, the potential energy reaches zero in the center of the strip, and becomes negative in higher fields. Above $B_{S}$, the vortices can exist in the film in a stable state \cite{Likharev}. However, the vortices still may not be able to enter (or exit) the strip because of the potential barriers at the edges of the strip. These barriers form due to the screenning currents around the edges \cite{Bean, Clem, Edges, Plourde}, and will prevent the entry of vortices until a higher magnetic field $B_{E}$ is reached, at which the potential barriers at the edges disappear.  
The effect of the applied current on the potential energy landscape accross the strip or a nanowire at a magnetic field $B_{S}$ is shown in Fig. 4. b. The potential is tilted by the applied current, effectively decreasing the barriers. For large enough currents, both the entry and the exit barrier disappear and the vortices can cross the wire, at which point the sample enters a resistive state. 

\begin{figure}
  \includegraphics[width=8 cm]{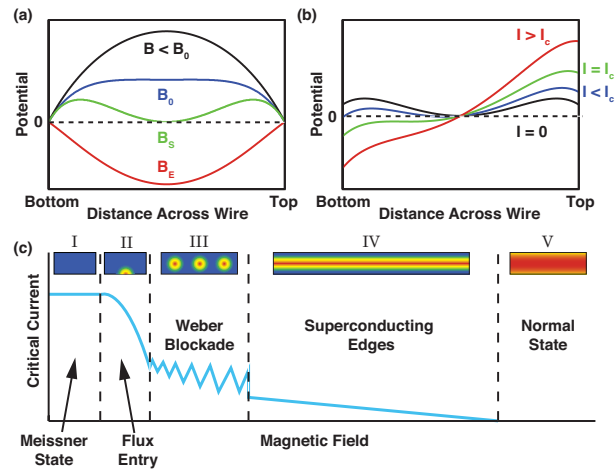}
  \caption{{\bf(a)} Potential energy of a vortex as a function of its position accross the width of the wire for different magnetic fields, with no applied current \cite{Weber}.
  {\bf(b)} Potential energy of a vortex as a function of its position along the width of the wire for different applied currents, in a magnetic field between $B_{0}$ and $B_{S}$.
    {\bf(c)} A schematic of different vortex regimes in the nanowire as a function of current and magnetic field. The blue color denotes the superconducting regions and the red color corresponds to the normal regions.}
  \label{4}
\end{figure}

The overall decrease of the critical current with increasing magnetic field is qualitatively well described by the Ginzburg-Landau model, apart from the oscillations observed in the middle range of magnetic fields. In order to explain these oscillations, we need to also consider the length of our samples. Specifically, if the nanowires are short enough that the vortices can be treated as discrete entities, than we may expect to the critical current to increase and decrease as each vortex enters the nanowire. This type of behavior was predicted by Vodolazov\cite{Vodolazov}, who studied the entry of vortices in narrow strips in the framework of the Ginzburg-Landau model. In the regime between $B_{0}$ and $B_{E}$, the vortices can exist in the wire, but there are energy barriers on both edges of the nanowire. In this regime, the onset of resistance occurs when vortices can both enter and exit the wire. Vodolazov calculated the potential energies for entry and exit of vortices, and found that both energies show saw-tooth oscillations due to entry (or exit) of a single vortex when the length of the wire is 40$\xi$ (for comparison, the length of our sample shown in Fig. 3 is 55$\xi$). The larger of these two barriers will be the bottleneck for vortex conduction, and will therefore determine $I_{C}$. Because of the finite length of the nanowire, the distance between vortices changes discontinuously when each vortex enters the nanowire, causing discontinuous jumps in the entry and exit barriers. For a fixed number of vortices in the nanowire, the increasing magnetic field will cause the exit barrier to increase and the entry barrier to decrease, until the next vortex enters the nanowire \cite{Vodolazov}. The observed periodicity corresponds to $\Phi_{0}$/2$\pi$Lw, where L is the length and w is the width of the wire.

These discrete changes in the potential barriers for entry and exit of vortices will directly affect the measured critical current. In the regime in which vortices can enter the nanowire, the critical current is determined by the sum of the applied current, vortex current and the screening currents generated in the superconducting regions. After the first vortex enters the nanowire, the increasing magnetic field will cause the nanowire to generate screening currents in order to expell any additional flux. In this regime, the critical current will be determined by the exit barrier - the onset of resistance will occur when the vortex can exit the nanowire. The exit barrier increases with magnetic field \cite{Vodolazov}, causing a corrresponding linear increase in the critical current, as found by Maximova \cite{Maksimova}. This will continue until the additional flux reaches one half of a flux quantum, after which the screening currents will change direction in order to generate additional flux to complete a full flux quantum. In ths regime, the critical current will be determined by the entry barrier, with the onset of resistance occuring when another vortex can enter the nanowire. This would result in linearly decreasing critical current \cite{Maksimova} (even though Maksimova assumed a continuous linear vortex density, which does not apply for our short wires with discrete vortices, it should be safe to extend her expressions for the magnetic field dependence of the current required for the entry and exit of the first vortex to the subsequent discrete vortices).

The full picture is then shown in Fig. 4. c., which gives the general behavior of the critical current as a function of magnetic field, observed in all our narrow and short nanowires, along with the schematics of the superconducting wavefuntion inside the nanowire in each regime. Region I is dominated by the Meissner effect, where the magnetic field is expelled from the nanowire. Region II marks the onset of the mixed state, where the magnetic field starts to partially penetrate the sample. Region III is the Weber blockade regime, where each oscillation signifies an addition of a single vortex to the nanowire. The vortices are arranged in a single row, as shown in the schematic, and the spacing between the vortex cores is generally larger than the width of the strip \cite{Vodolazov}. As the magnetic field is increased further, the potential barriers at the edges decrease, the vortex row becomes denser and vortices start to interact, gradually merging into a superconducting channel in the center of the strip \cite{Palacios}. Region IV shows a slow linear decrease in the critical current until the magnetic field is large enough to destroy the surface superconductivity at the edges of the nanowires and the sample enters the normal state, marked by Region V. The drawings of the vortices in the nanowire are shown in the case without bias current. The bias current would shift the vortices towards one edge of the nanowire (top or bottom on the images, depending on the direction of the magnetic field).

An equivalent description of the Weber blockade regime in Region III can be obtained by considering only the vortex degrees of freedom. In this regime, vortices can exist in the nanowire in a stable state, but there are energy barriers to both entry and exit of vortices. The nanowire is short enough that the states containing different numbers of vortices have distinct energies. Using the vortex-charge duality, this system can be viewed as a vortex analog of a quantum dot. The magnetic field is then the analog of the gate voltage, as it tunes the chemical potential (the number of vortices in the nanowire). The applied current exerts a force on the vortices, causing them to move towards one edge of the nanowire and eventually leave the nanowire. In that sense, the current bias in the Weber dot is the analog of the bias voltage in the Coulomb dots. This type of a system was considered by Pekker \textit{et al.} \cite{Weber}, who obtained "Weber diamonds" of zero resistance as a function of magnetic field and bias current. In the dual picture, zero resistance corresponds to zero conductance for vortices, in analogy with the Coulomb diamonds in quantum dots, inside which the electron conductance iz zero. The calculations of Pekker \textit{et al.} \cite{Weber} were carried out for samples with thickness of the order of $\xi$, width of <5$\xi$ and length between 10$\xi$ and 100$\xi$, which corresponds to the dimensions of our samples. In our samples, if we remove the background slope, the oscillations shown in Fig. 3. c. represent the top half of the diamond pattern (the bottom half is observed for the bias current of opposite polarity). As in the conventional picture, the up slope is determined by the exit of vortices (analogous to one of the leads in the Coulomb blockade), and the down slope by the entry (analogous to the other lead in the Coulomb blockade). 

We conclude that narrow and short superconducting nanowires can exist in a ground state which contains a fixed number of vortices, precisely tunable by applied magnetic field. While this may cause resistance fluctuations and noise in devices which involve superconducting nanowires \cite{Shahar}, it may also be used as an advantage in vortex-based devices.

\vskip 0.2in

The authors would like to thank D. Pekker and V. Oganesyan for useful comments and suggestions. This work was supported by NSF Grant No. DMR-1106167. N. M. would like to thank the Aspen Center for Physics (through the NSF Grant No. PHYS-1066293) for 
hospitality. 
%

\bibliography{WeberBibliography}

\end{document}